\documentclass[%
 reprint,
superscriptaddress,
 amsmath,amssymb,
 aps,
prb,
]{revtex4-2}

\usepackage{graphicx}
\usepackage{dcolumn}
\usepackage{bm}
\usepackage{hyperref}

\usepackage{mathtools,amssymb}
\usepackage{xcolor}
\usepackage{soul}
\begin{document}

\preprint{APS/123-QED}

\title{Stability limits in two-band superconductor rings}

\author{Leonardo Rodrigues Cadorim}
\affiliation{Departamento de F\'isica, Faculdade de Ci\^encias, 
Universidade Estadual Paulista (UNESP), Caixa Postal 473, 
17033-360, Bauru-SP, Brazil}
\affiliation{Department of Physics, University of Antwerp, Groenenborgerlaan 171, B-2020 Antwerp, Belgium}
\author{Edson Sardella}
\affiliation{Departamento de F\'isica, Faculdade de Ci\^encias, 
Universidade Estadual Paulista (UNESP), Caixa Postal 473, 
17033-360, Bauru-SP, Brazil}

\author{Daniel Domínguez}
\affiliation{Centro Atómico Bariloche and Instituto Balseiro,
8400 San Carlos de Bariloche, Argentina}

\author{Jorge Berger}
\affiliation{Department of Physics, Braude College, 2161002 Karmiel, Israel}





\date{\today}

\begin{abstract}
This study explores transitions between states with different winding number in 
two-band superconducting rings. From the 
time-dependent Ginzburg-Landau (TDGL) equations for 
two-component superconductors, we apply linear 
instability theory and develop a semi-analytical 
method that provides the critical flux for 
phase-slip occurrence. The developed method was 
applied to investigate how the critical flux depends 
on physical properties, such as band 
parameters and temperature. Finally, we 
show the possible existence of a soliton state, 
in which the phase winding number is different in each condensate.
\end{abstract}

\maketitle


\section{\label{sec:level1}Introduction}

The existence of different overlapping sheets at the Fermi surface 
of a superconductor gives rise to the so called multiband 
superconductivity, characterized by more than one condensate. \cite{xi2008,tanaka2015} 
Since several of the novel superconducting materials that emerged 
in the last years are multiband superconductors, this topic has 
seen a revival of its interest in the literature, after beeing 
discussed for the first time in the late fifties. \cite{suhl1959} 
As examples of such materials, we can mention MgB$_2$, \cite{nagamatsu2001,bouquet2001} 
iron pnictides, \cite{paglione2010} NbSe$_2$, \cite{yokoya2001,rossnagel2001} 
iron arsenides, \cite{rotter2008} among others.

The presence of different condensates has also rendered multiband 
superconductors of great interest due to their magnetic properties. 
One of them is the possibility of non-monotonic vortex-vortex 
interaction, with short-range repulsion and long-range attraction, 
arising through the competition between the coherence length of 
each condensate. \cite{babaev2005} Such interaction leads to 
the formation of vortex clusters, separated by Meissner state 
regions in bulk specimens. \cite{babaev2005} This vortex matter 
behavior is known as type-$1.5$ superconductivity \cite{moshchalkov2009} 
and has raised a great amount of discussion. 
\cite{babaev2010,carlstrom2011,kogan2011,babaev2012,shanenko2011,chaves2011}

Another phenomenon observed in multiband superconductors is 
the possibility of vortices carrying a non-integer multiple 
of the flux quanta. \cite{babaev2002,babaev2009} Such objects 
can occur either when the phase winding of different condensates 
are not equal or when the phase singularities of each condensate 
occur at different points inside the superconductor. Although 
the existence of fractional vortices in bulk superconductors 
s thermodynamically prohibited due to their divergent energy, 
it is known that their formation in mesoscopic samples is 
possible. \cite{chibotaru2010,geurts2010,pina2012} It was 
shown that fractional vortices can also be stabilized near 
the surface \cite{silaev2011} and that such configuration also
has an effect in the magnetization curve of multiband 
superconductors. \cite{dasilva2014}

It was also shown that the multiband superconductors 
have interesting properties in the presence of an 
external current. For instance, two-band superconductors 
can display a state where the phase difference between 
their two condensates are neither locked at $0$ or $\pi$, 
the so called phase soliton state. \cite{tanaka2001,kuplevakhsky2011,yerin2021} 
In $1D$ two-band systems, it was shown that such 
states can emerge from the application of a current 
to the superconductor, \cite{gurevich2003,gurevich2006} 
with the detectable appearance of oscillations of 
the critical current with the length of the system, 
provided that the interband coupling is small. 
\cite{marychev2018} In two-band systems with higher 
dimensions, the possibility of vortex formation and 
its subsequent motion give origin to nonequilibrium 
phase textures, which significantly affects the 
dissipation process. \cite{polo2017}
Recently, the formation of phase-slips in quasi-one dimensional two-band superconductors wires was studied and the time evolution of phase solitons and the conditions for their existence is such scenario were systematically described.

In the case of a superconducting loop that encloses an applied flux, the single-valuedness of the order parameters ensures that the fluxoid is quantized. The passage between different fluxoid numbers necessarily involves the vanishing of an order parameter at some instant and position, permitting the occurrence of a phase slip. Such a system was extensively studied for single band superconductors \cite{vodolazov2003,vodolazov2002,kenawy2020,lu2009,berger2003,tarlie1998,karttunen2002}, with prospects of utilization in a number of 
applications, such as a superconducting qubit and a series of electronic devices \cite{mooij2005,mooij2006,astafiev2012,gumucs2023,de2018,ligato2022,murphy2017,hongisto2012,cheng2018,di2021}. Of particular interest to the present work is the deterministic phase-slip, which occurs when the energy barrier separating two flux states vanishes. \cite{tarlie1998,petkovic2016}
For single-band superconductors, analytical expressions for the conditions to the occurrence of a deterministic phase-slip have been found combining linear instability theory and Ginzburg-Landau formalism. \cite{tarlie1998,karttunen2002}
In this work we use linear stability theory together with the Ginzburg-Landau formalism for two-band superconductors and derive a semi-analytical method to obtain the critical flux for the transition between two fluxoid states in a superconducting loop. Furthermore, we show that our analytical results 
are in agreement with the full Ginzburg-Landau 
theory and investigate how the system behavior depends 
on parameters such as the ring radius, the ratio 
between the diffusion coefficients of each band, and the temperature. In Refs.~\onlinecite{tarlie1998,karttunen2002}, the stability of single-band superconductors was studied under the action of an external field varied at a finite rate. In the present work, when investigating the stability of different fluxoid states in two-band superconductors, the flux is either kept fixed or changed quasistatically.

The structure of this manuscript is as follows. 
In Sec.~\ref{sec:level2} we introduce the 
time-dependent Ginzburg-Landau equations for 
two-band superconductors. In Sec.~\ref{sec:level3} 
we detail how linear stability theory is used 
to obtain stability limits for our system 
and compare them with fully numerical results. 
We apply our method under different physical 
conditions in Sec.~\ref{sec:level4}. Finally, 
we present our concluding remarks in Sec.~\ref{sec:level5}.

\section{\label{sec:level2}Theoretical Formalism}

The time dependent Ginzburg-Landau equations for a two-band system, in 
dimensionless units, are given as follows 
\cite{vargunin2020}:

\begin{eqnarray}
    \left (\frac{\partial}{\partial t}+i\varphi \right )\Delta_1 &=& -\left (-i\mbox{\boldmath $\nabla$}-{\bf A}\right )^2\Delta_1 \nonumber \\
    & &+\left(\chi_1-|\Delta_1|^2\right)\Delta_1+\gamma\Delta_2\;, \label{eqn:eqn1}\\
    \left (\frac{\partial}{\partial t}+i\varphi \right )\Delta_2 &=& -\frac{D_2}{D_1}\left (-i\mbox{\boldmath $\nabla$}-{\bf A}\right )^2\Delta_2 \nonumber \\
    & &+\left(\chi_2-|\Delta_2|^2\right)\Delta_2+\frac{\eta_1}{\eta_2}\gamma\Delta_1\;, \label{eqn:eqn2}
\end{eqnarray}
and Ampère's law is:
\begin{equation}
    \frac{1}{u}\left[1+\frac{\eta_2}{\eta_1}\frac{D_2}{D_1}\right] \left (\frac{\partial {\bf A}}{\partial t}+\mbox{\boldmath $\nabla$} \varphi \right ) = {\bf J}_s-\kappa_1^2\mbox{\boldmath $\nabla$}\times {\bf h}\;,  \label{eqn:eqn3}
\end{equation}
where the supercurrent density is given by:
\begin{equation}
    {\bf J}_s = \sum\limits_{j=1}^{2} \frac{\eta_j}{\eta_1}\frac{D_j}{D_1} \textrm{Re}\left[\Delta_j^*\left (-i\mbox{\boldmath $\nabla$}-{\bf A}\right )\Delta_j\right]\;. \label{eqn:eqn4}
\end{equation}

In the equations above, $D_j$ and $\eta_j$
are the diffusion coefficient and the partial density of states
of band $j$; $\chi_j$
is defined as $\chi_j = (T_{cj}-T)/T_c$,
where the ratio between the critical temperature of the $j$
band and the critical temperature
of the coupled system ($T_{cj}/T_c$)
is defined in Ref.~\onlinecite{kogan2011};
$\gamma$ is defined as $\gamma = \lambda_{12}/(\eta_1\delta)$,
where $\delta = \lambda_{11}\lambda_{22} -\lambda_{12}^2$ is the determinant of the coupling constant
matrix with components $\lambda_{ij}$.
Finally, $\kappa_1$ is defined as the ratio between
the coherence length at zero temperature $\xi_1 = \sqrt{\pi\hbar D_1/8T_c}$
and the penetration length at zero temperature
$\lambda_1 = 
\sqrt{(1/N(0)\eta_jD_1)(7\zeta(3)\hbar c^2/32\pi^4e^2T_c)}$
of the first band.
The order parameters $\Delta_j$
are in units of $\sqrt{8\pi^2T_c^2/7\zeta(3)}$;
lengths are expressed in units of the coherence length
of the first band $\xi_1$; the magnetic field and the vector
potential are given in units of $H_{c2}^{(1)}$
and $H_{c2}^{(1)}\xi_1$, respectively,
where $H_{c2}^{(1)} = \Phi_0/(2\pi \xi_1^2)$,
with $\Phi_0 = hc/2e$ being the magnetic flux quantum unit;
the scalar potential is in units of
$\varphi_0 = H_{c2}^{(1)}D_1/c$;
the current densities in units of
$j_0 = 4eN(0)\eta_1\pi^3T_cD_1/(7\zeta(3)\xi_1)$;
and time in units of $t_0 = \xi_1^2/D_1$.
In Eq.~(\ref{eqn:eqn3}), we have $u = 5.79$,
as in the case of single-band superconductors.

We have chosen the microscopic parameters of our two-band
superconductor as follows.
The coupling constants are given by $\lambda_{11} = 2.0$,
$\lambda_{22} = 1.03$ and $\lambda_{12} = 0.005$.
The partial density of states of band $1$ is $\eta_1 = 0.355$
($\eta_2 = 1-\eta_1$).
These parameters correspond to a critical temperature
of each band given by $T_{c1} = 0.9997 T_c$
and $T_{c2} = 0.903 T_c$.
The ratio $D_1/D_2$ is left as a free parameter for our subsequent analysis. Our problem consists of finding the local minimum of the system energy and then investigate at which critical flux this minimum becomes a saddle point \textit{i.e.} the flux at which it is no longer stable. To do so, we use the TDGL equations.

As derived in Ref.~\onlinecite{vargunin2020},
the TDGL equations above are strictly valid only for gapless
superconducting states.
However, since we are using TDGL as a mathematical tool to find a local minimum, our procedure is valid as long as the Ginzburg-Landau expression for the free energy remains valid. As a by-product, we will obtain a qualitative description of the transient passage between states and the phase-slip that occurs during this passage.

In what follows, we consider a superconducting ring
of radius $R$ placed in an homogeneous applied field
in such a manner that the ring in the normal state would enclose an amount of flux $\Phi$.
We consider that the ring width and thickness are much smaller
than all characteristic lengths of the system
(the coherence lengths of each band, the penetration length
of the superconductor and $R$), thus the ring is effectively one-dimensional. This implies that the Ampère law (\ref{eqn:eqn3})
does not need to be solved, the vector potential can be taken uniform in the entire ring and is given by 
$\bm{A} = \Phi/(2\pi R)$.

When numerically solving the time dependent Ginzburg-Landau
equations, we applied the link-variable method 
\cite{gropp1996} which guarantees gauge invariance throughout
the numerical procedure. In order to simulate the ring
geometry for the one-dimensional system,
we use periodic boundary conditions $\Delta_j(0) = \Delta_j(L)$,
$\varphi(0) = \varphi(L)$, where $L$ is the ring length
$L = 2\pi R$. The use of periodic boundary condition gives
rise to a numerical problem, since
every single point of the grid is equivalent.
In such scenario, nonuniform states would never be encountered, since there is no preferred
position for the nucleation of this configuration.
To overcome this difficulty, we 
randomly insert inhomogeneous
spots in the ring by locally changing the temperature
by a physically negligible amount in the order of $10^{-5} T_c$.
The location of these spots and the specific value
of the temperature increment do not alter the 
results presented here.

\section{\label{sec:level3}Linear Stability Theory}

In this section, we apply linear stability theory
to the problem of passage between states with different winding number in a two-band
superconductor. To our knowledge, this method has been used in the past only for single-band superconductors.

Our starting point is the first Ginzburg-Landau 
equation in the one-dimensional form:

\begin{eqnarray}
    \left (\frac{\partial}{\partial t}+i\varphi \right )\Delta_1 &=& -\left (-i\frac{\partial}{\partial x}-A\right )^2\Delta_1 \nonumber \\
    & &+\left(\chi_1-|\Delta_1|^2\right)\Delta_1+\gamma\Delta_2\;, \label{eqn:eqn5}\\
    \left (\frac{\partial}{\partial t}+i\varphi \right )\Delta_2 &=& -\frac{D_2}{D_1}\left (-i\frac{\partial}{\partial x}-A\right )^2\Delta_2 \nonumber \\
    & &+\left(\chi_2-|\Delta_2|^2\right)\Delta_2+\frac{\eta_1}{\eta_2}\gamma\Delta_1\;, \label{eqn:eqn6}
\end{eqnarray}
where $x$ represents an arbitrary arc-length.

We then write each order parameter as a stationary
term plus a small perturbation
$\Delta_j = \Delta_j^0+\delta\Delta_j$,
with the stationary unperturbed term given by
$\Delta_j^0 = a_j e^{in_ix/R}$, where $n_i$
is the winding number of the initial state.
At present, we have limited this analysis to the case in which both bands have the same winding number. However, in our later discussion on numerically solving 
Eqs.~(\ref{eqn:eqn1})-(\ref{eqn:eqn3}), we will 
show that after reaching the critical flux the winding numbers for each of the bands will not always be the same and the change in winding number will not always be unity.
The values of $a_j$
can be found from the procedure employed in
Ref.~\cite{chaves2011}; for completeness, we reproduce
it below.

Substituting $\Delta_j^0$ in Eqs.~(\ref{eqn:eqn5}) 
and (\ref{eqn:eqn6}), with $\varphi = 0$, we obtain the following set of equations:

\begin{eqnarray}
    &-&\left (n_i/R-A\right )^2a_1+\left(\chi_1-a_1^2\right)a_1+\gamma a_2 = 0\;, \label{eqn:eqn7}\\
    &-&\frac{D_2}{D_1}\left (n_i/R-A\right )^2a_2+\left(\chi_2-a_2^2\right)a_2+\frac{\eta_1}{\eta_2}\gamma a_1 = 0\;. \label{eqn:eqn8}
\end{eqnarray}

Defining $\rho = a_1/a_2$, a combination of 
Eqs.~(\ref{eqn:eqn7}) and (\ref{eqn:eqn8}) produces 
the following polynomial equation for $\rho$:

\begin{eqnarray}
    \frac{\eta_1}{\eta_2}\gamma\rho^4 &+&\left(\chi_2-\frac{D_2}{D_1}(n_i/R-A)^2\right)\rho^3 \nonumber \\
    &-&\left(\chi_1-(n_i/R-A)^2\right)\rho-\gamma = 0\;, \label{eqn:eqn9}
\end{eqnarray}

After $\rho$ is known, $a_1$ and $a_2$ are given by:

\begin{eqnarray}
    a_1 & = & \sqrt{\gamma/\rho+\chi_1-(n_i/R-A)^2} \nonumber \\
    a_2 & = & \sqrt{\frac{\eta_1}{\eta_2}\gamma\rho+\chi_2-\frac{D_2}{D_1}(n_i/R-A)^2}\;.
\end{eqnarray}

To write down the perturbation, we follow 
the same procedures used for single-band systems. Since single-valuedness of the order parameters implies that the perturbation has to be periodic, we express it as a Fourier expansion:
\begin{equation}
    \delta\Delta_j = \sum_{n_f}[\hat{a}_{n_f}^je^{in_fx/R}+\hat{a}_{-n_f}^{j*}e^{i(2n_i-n_f)x/R}]e^{\lambda_{n_f} t}\;, \label{eqn:eqn11b}
\end{equation}
here $\lambda_{n_f}$ is real and ${n_f}$ is the winding number of the final state.
We remind that $n_f$ is also set to be the same for both bands.
As we can see, the parameter $\lambda_{n_f}$ determines 
the stability of the initial state against the $n_f$ mode of the perturbation, 
since it decreases exponentially if $\lambda_{n_f}$ is 
negative and grows exponentially otherwise. To determine 
the value of $\lambda_{n_f}$ for a given applied flux, 
we need to substitute $\Delta_j = \Delta_j^0+\delta\Delta_j$ 
in Eqs.~(\ref{eqn:eqn5}) and (\ref{eqn:eqn6}) and linearize the 
resulting expressions with respect to $\delta\Delta_j$. 
Making use of Eqs.~(\ref{eqn:eqn7}) and (\ref{eqn:eqn8}), 
we find:

\begin{eqnarray}
    &&\frac{\partial\delta\Delta_1}{\partial t}+i\varphi\Delta_1^0 = -\left (-i\frac{\partial}{\partial x}-A\right )^2\delta\Delta_1 \nonumber \\
    & &+\left(\chi_1-2a_1^2\right)\delta\Delta_1-a_1^2e^{2in_ix/R}\delta\Delta_1^*+\gamma\delta\Delta_2\;, \label{eqn:eqn10}\\
    &&\frac{\partial\delta\Delta_2}{\partial t}+i\varphi\Delta_2^0 = -\frac{D_2}{D_1}\left (-i\frac{\partial}{\partial x}-A\right )^2\delta\Delta_2 \nonumber \\
    & &+\left(\chi_2-2a_2^2\right)\delta\Delta_2-a_2^2e^{2in_ix/R}\delta\Delta_2^*+\frac{\eta_1}{\eta_2}\gamma\delta\Delta_1\;. \label{eqn:eqn11}
\end{eqnarray}

The scalar potential can obtained from the expression 
$\sigma_n \bm{\nabla}\varphi = -\bm{J}_n$, where 
$\sigma_n = (1/u)\left(1+\eta_2D_2/(\eta_1D_1)\right)$ 
is the normal conductivity. Using charge conservation 
and assuming there is no charge accumulation, we have 
$\bm{\nabla}\cdot(\bm{J}_n+\bm{J}_s) = 0$, from which the 
equation for the scalar potential in our one-dimensional 
system can be written as:

\begin{equation}
    \sigma_n\frac{\partial^2\varphi}{\partial x^2} = \frac{\partial J_s}{\partial x}\;.    
\end{equation}

Since the stability limits of our system depend on its energy landscape but not on its dynamical parameter, we can take the limit $\sigma_n \rightarrow \infty$, which allows us to take $\varphi = 0$.
With this result and using Eqs.~(\ref{eqn:eqn10})-(\ref{eqn:eqn11}), we are 
now able to write down an eigenvalue-eigenvector 
equation for each $\lambda_{n_f}$, which is given by:

\begin{gather}
    \begin{bmatrix}
        M_{11} & M_{12} & M_{13} & M_{14} \\
        M_{21} & M_{22} & M_{23} & M_{24} \\
        M_{31} & M_{32} & M_{33} & M_{34} \\
        M_{41} & M_{42} & M_{43} & M_{44}
        \end{bmatrix}
        \begin{bmatrix}
        \hat{a}_{n_f}^1 \\
        \hat{a}_{-n_f}^1 \\
        \hat{a}_{-n_f}^2 \\
        \hat{a}_{n_f}^2
        \end{bmatrix}
        =
        \lambda_{n_f}
        \begin{bmatrix}
        \hat{a}_{n_f}^1 \\
        \hat{a}_{-n_f}^{1*} \\
        \hat{a}_{-n_f}^{2*} \\
        \hat{a}_{n_f}^2
    \end{bmatrix}
    \label{eqn:eqn14}
\end{gather}
where the coefficients $M_{ij}$ are presented separately 
in Appendix \ref{sec:appA}. A detailed and cumbersome analysis of the matrix $M$ can show that its determinant ($det(M)$) depends on the applied flux only through $(n_i-A)^2$. Also, since the determinant of a matrix equals the product of its eigenvalues, we have that $det(M) = 0$ at a critical flux. Combined, these two results mean that the critical points for the transitions $n \rightarrow n+1$ and $n+1 \rightarrow n+2$ are separated by $\Phi_0$, as physically expected. In addition, we conclude that the upper and the lower stability limits of a state with winding number $n$ are equidistant from $n\Phi_0$.

\begin{figure}[t!]
    \centering
    \includegraphics[width=0.48\textwidth]{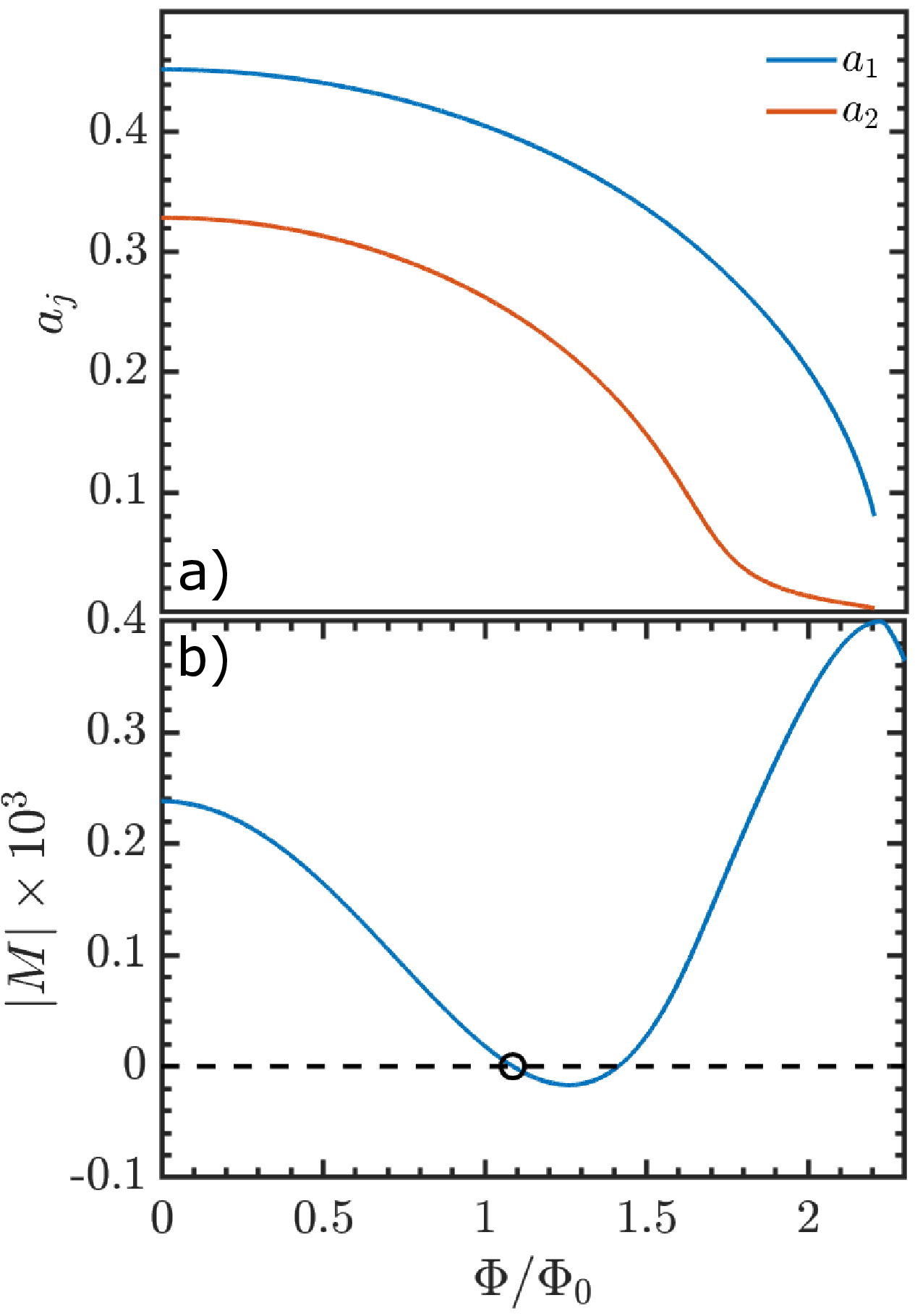}
    \caption{(Color online) The upper panel shows the 
    values of $a_1$ and $a_2$ for $n_i=0$ as functions of the applied 
    flux. The lower panel presents $det(M)$ as a function of $\Phi$; the 
    black circle indicates the critical flux $\Phi = 1.087 \Phi_0$, 
    above which the state with winding number $0$ is unstable.}
    \label{fig:fig2}
\end{figure}

Furthermore, we can now clearly see the reason for assuming $n_f$ to be 
equal in both bands. Since we solve Eq.~(\ref{eqn:eqn14})
for each $n_f$ separately,
$n_f$ must be the same in the two bands. Nevertheless, 
it will be shown that the results from our method are 
still valid even if the winding number of each band
is not the same at the final state \textit{i.e.}
even if the final state presents a phase soliton.  

At this point, we introduce a compromise into our 
previously entirely analytical procedure by adopting 
numerical routines moving forward. This is done because, in order 
to get the final value of $\Phi_c$, one needs first 
to solve the fourth order polynomial Eq.~(\ref{eqn:eqn9}) for 
$\rho$ and subsequently for $a_1$ and $a_2$. With this result, $det(M)$ must then be calculated for different values of $\Phi$, until $det(M) = 0$ is found. In principle, while it is possible to carry on this procedure analytically, the laborious 
work required to do so is not worth, since we can reach 
the same results with an almost negligible computational 
effort. In addition, given the complexity of these equations, 
we probably would obtain intractable results.
 
To exemplify our method, 
we apply it to the problem of finding the critical flux
at which the Meissner state of a two-band superconducting
ring with radius $R = 5 \xi_1$ and $D_1/D_2 = 1.00$
is unstable against a perturbation with winding number $n_f$.
As is the case for single-band superconductors,
the first mode to become unstable corresponds to $n_f = 1$.
Therefore, in the following we solve Eq.~(\ref{eqn:eqn14})
with $n_i = 0$ and $n_f = 1$.

\begin{figure}[t!]
    \centering
    \includegraphics[width=0.48\textwidth]{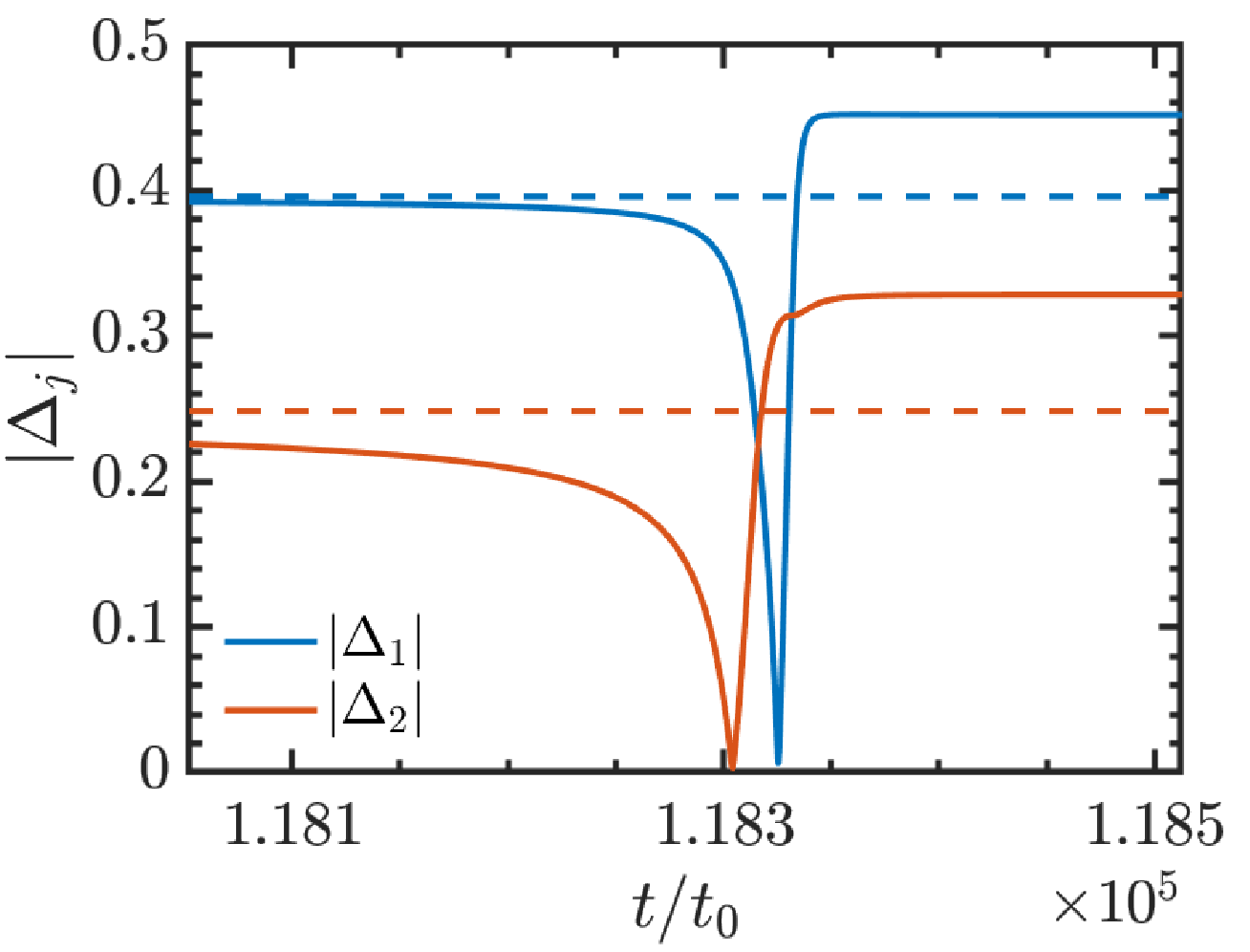}
    \caption{(Color online) Time evolution of the minimum values of the order parameters of a state initially prepared with winding number 0, close to the limit of stability, $\Phi_c=1.087\Phi_0$. $R = 5\xi_1$, $D_1=D_2$, $T=0.8T_c$. Solid and dashed lines represent the evolution for $\Phi = \Phi_c+0.001\Phi_0$ and $\Phi = \Phi_c-0.001\Phi_0$, respectively.}
    \label{fig:fig3}
\end{figure}

Panel $a)$ of 
Fig.~\ref{fig:fig2} presents the values of $a_1$ and 
$a_2$ as functions of the applied flux.
The region of high applied flux shows an important feature
for our future discussions. As can be seen,
there is a pronounced change in the behavior
of the curve corresponding to $a_2$,
which acquires an elongated tail. 
As can be easily visualized from Eqs.~(\ref{eqn:eqn7})-(\ref{eqn:eqn8}), 
this occurs because the applied flux reaches a 
value higher than the critical flux at which the 
second band goes to the normal state in the decoupled regime.
In other to avoid confusion between this critical flux
and the critical flux at which the initial state becomes unstable, hereafter we refer to the former as the "upper critical
flux of the second band".
Above this flux, the second band
stays superconducting due only to coupling
with the first band. Eq.~(\ref{eqn:eqn8})
and the expression of the vector potential
show that the upper critical flux of 
the second band increases with $D_1/D_2$
and with the radius of the ring $R$.

The panel $b)$ of Fig.~\ref{fig:fig2} shows $det(M)$ as a function of $\Phi$, with the black 
circle indicating the value of flux ($\Phi_c = 1.087 \Phi_0$) at which the first eigenvalue of \label{eqn:eqn14} 
changes its sign. This is 
the critical flux value, above which the system 
is unstable against decay into a state with winding number $1$.

While our semi-analytical method allows us to calculate
the flux at which a state becomes unstable,
the full numerical solution
of Eqs.~(\ref{eqn:eqn1})-(\ref{eqn:eqn3})
gives us the time evolution of both order parameters
for a given applied flux.
Fig.~\ref{fig:fig3} exemplifies the above analysis by showing the time evolution of the minimum values of each order parameter for a system with the same parameters described in Fig.~\ref{fig:fig2}. Here, dashed lines represent a case with $\Phi < \Phi_c$, while solid lines show the order parameter evolution for $\Phi > \Phi_c$. For $\Phi < \Phi_c$,
the system just stays the whole time in the Meissner state,
without major modifications. However, if $\Phi > \Phi_c$, the 
system initially sits practically unperturbed in the Meissner state. 
Eventually, the perturbation becomes noticeable, and the 
system transits to a state with winding number greater than $0$. As expected, the occurrence of a phase-slip is not simultaneous in both bands, with the time interval between between them depending on $D_1/D_2$, the applied flux and temperature. Although both phase slips are not simultaneous, they do occur at the same place, i.e. they are a case of time-connected phase-slips. \cite{dominguez2024} After this couple of phase-slips, the system stabilizes at a new state, with winding number 1.

\begin{figure}[t!]
    \centering
    \includegraphics[width=0.48\textwidth]{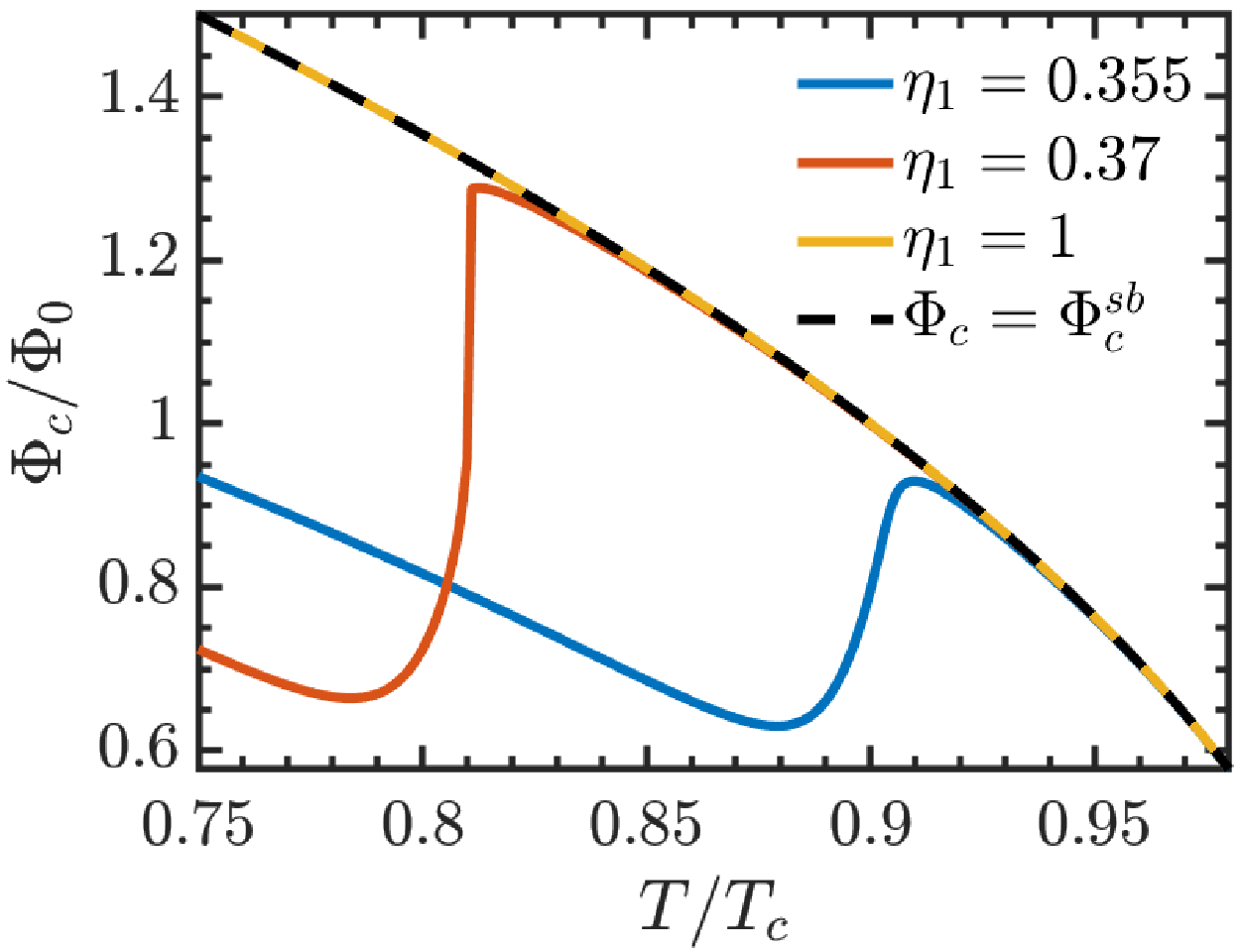}
    \caption{(Color online) Critical flux for the emergence of a phase slip as a function of temperature for different values of $\eta_1$. The black dashed curve shows the single-band limit $\Phi^{sb}_c = \sqrt{(1-T)R^2+1/2}/\sqrt{3}$.}
    \label{fig:fig3b}
\end{figure}

As a test of our semi-analytical method we compare the results that it yields with the well known expression for the single-band limit \cite{tuckerman1990,tarlie1998,karttunen2002} $\Phi^{sb}_c = \sqrt{(1-T)R^2+1/2}/\sqrt{3}$. Fig.~\ref{fig:fig3b} shows this comparison, by presenting the critical flux $\Phi_c$ as a function of the temperature for different values of $\eta_1$. As we can see, as $\eta_1$ increases and the importance of the second band to the physical behavior of the system decreases, the curve approaches the single-band limit, represented by the black dashed curve. In particular, for $\eta_1 = 1$, both models coincide, supporting the correctness of our semi-analytical method.

\section{\label{sec:level4}Results and Discussion}
 
Having presented our semi-analytical 
method and an example of its application, let us 
now apply it to a specific study. We will 
investigate how the critical flux, above which 
the Meissner state first becomes unstable against decay into a state with winding number $1$, depends on the 
ratio between the diffusion coefficients in 
each band and on the bath temperature.

    \begin{figure}[t!]
        \centering
        \includegraphics[width=0.48\textwidth]{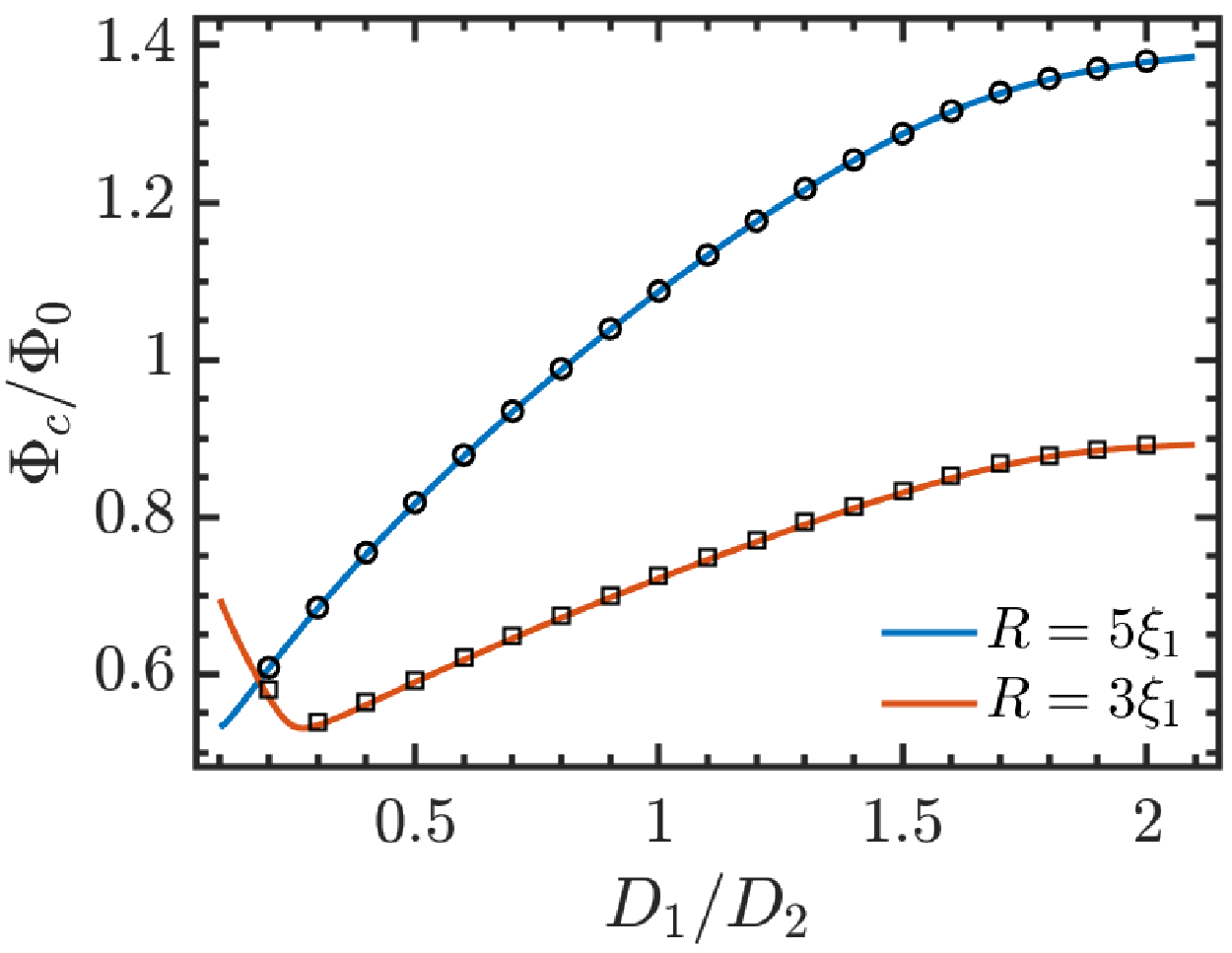}
        \caption{(Color online) Critical flux above which 
        the system becomes unstable 
        as a function of the ratio $D_1/D_2$. The blue (red) 
        line presents the results obtained from our method 
        developed in Section \ref{sec:level3} 
        for a ring with radio $R = 5 \xi_1$ ($R = 3 \xi_1$). 
        Black circles (squares) show the values of the 
        critical flux obtained numerically directly from the TDGL 
        Eqs.~(\ref{eqn:eqn5}) and (\ref{eqn:eqn6}).}
        \label{fig:fig4}
    \end{figure}

    \subsection{The critical flux dependence on $D_1/D_2$}

    We have already seen that our linear stability analysis is correct in the single-band limit, let us now compare its results with the direct numerical
    solution of Eqs.~(\ref{eqn:eqn1})-(\ref{eqn:eqn3}) when both bands are active. 
    In Fig.~\ref{fig:fig4}, we present the critical flux at
    which the two-band system first becomes unstable, as a function of the ratio of the
    diffusion coefficients $D_1/D_2$.
    The blue (red) line shows the critical flux obtained from our semi-analytical method for a ring with radius $R = 5\xi_1$ ($3\xi_1$), whereas the black circles (squares) indicate the critical flux obtained by solving numerically the TDGL equations for a ring with $R = 5\xi_1$ ($3\xi_1$).
    The black circles indicate 
    the critical flux obtained by numerically solving 
    the TDGL equations for the system with 
    $R = 5\xi_1$, and the black squares represent 
    the results for $R = 3\xi_1$. 
    As can be seen, the agreement between our 
    semi-analytical method and the numerical results fully 
    verifies the validity of our procedure. In fact, the 
    largest deviation between them is of the order of $10^{-3} \Phi_0$.
    
\begin{figure}[t!]
        \centering
        \includegraphics[width=0.45\textwidth]{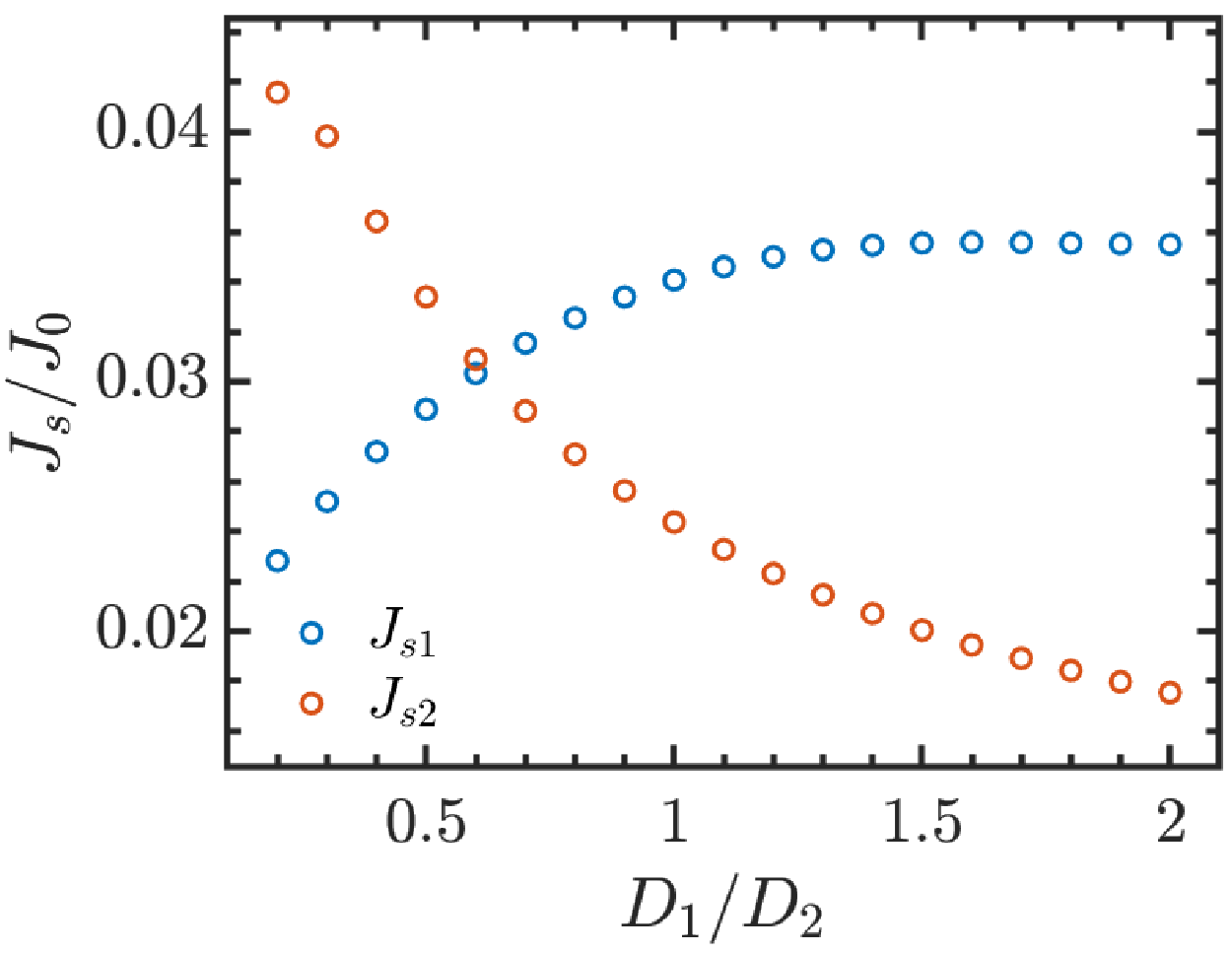}
        \caption{(Color online) Supercurrent 
        carried by the first (blue circles) and 
        second band (red circles) in the Meissner 
        state for an applied flux equals to the 
        critical flux of the given $D_1/D_2$ as 
        a function of $D_1/D_2$, for $R = 5 \xi_1$}
        \label{fig:fig5}
    \end{figure}

    Analyzing the curve corresponding to $R = 5 \xi_1$,
    one can see that the critical flux increases with $D_1/D_2$,
    reaching an asymptotic value for large diffusion coefficient ratio.
    This can be explained by the amount of current carried by 
    each condensate, which depends on the relative strength 
    of the bands and the values of $D_1$ and $D_2$. Since 
    all microscopic constants are held fixed and $D_1$ is 
    used in the definition of our units, the proportion of 
    current in each condensate is determined by the ratio 
    $D_1/D_2$. An increase in this ratio (decrease in 
    $D_2$) means a decrease in the amount of current 
    carried by the second condensate. This can be 
    seen in Fig.~\ref{fig:fig5}, where we show the 
    amount of current carried by each band 
    (calculated exactly at $\Phi_c(D_1/D_2)$) as 
    a function of $D_1/D_2$.
    
    As the current in the second band decreases,
    the phase-slip formation process 
    becomes more dependent
    on the dynamics of the first band. Since the first band
    has a higher critical temperature, the critical flux
    for the formation of a phase-slip increases,
    explaining the general behavior of the curve.
    If the ratio $D_1/D_2$ becomes too large,
    the dynamics of the system is determined solely
    by the first band and thus the critical flux
    tends to a constant value, which is also in agreement
    with our results.
    
    If we now examine the curve 
    corresponding to $R = 3 \xi_1$,
    we see that a special feature, 
    that is,
    a minimum in the critical flux is now present.
    Here,
    for small values of $D_1/D_2$, the critical flux 
    for phase-slip formation decreases, 
    contrary to increasing, as the diffusion 
    coefficient ratio increases.

    \begin{figure}[t!]
        \centering
        \includegraphics[width=0.48\textwidth]{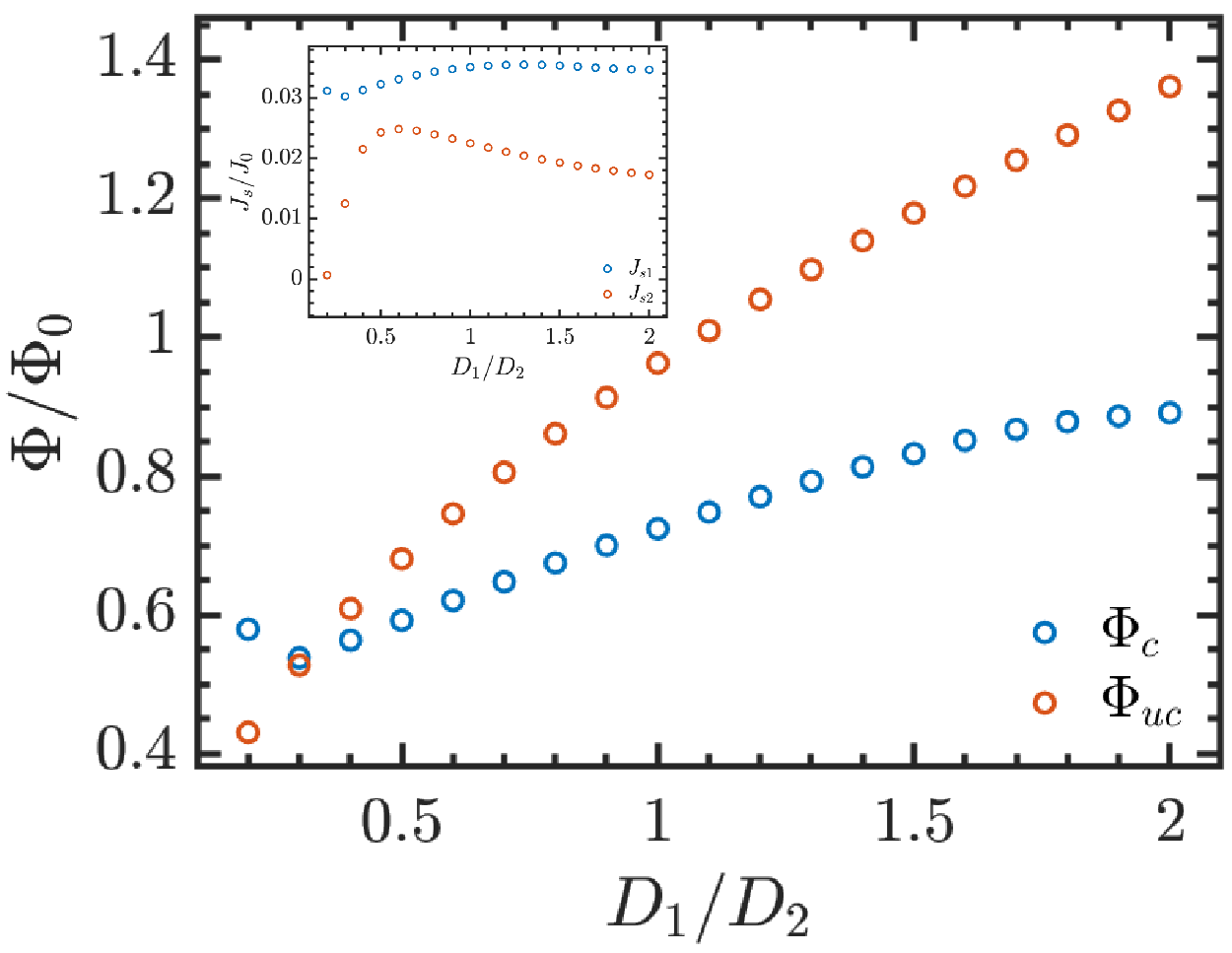}
        \caption{(Color online) The main panel 
        shows the critical flux for the formation 
        of a phase-slip (blue circles) and the 
        upper critical field of the second band 
        (red circles) as functions of $D_1/D_2$. 
        The inset presents the supercurrent carried 
        by the first (blue circles) and second 
        (red circles) band in the Meissner state, 
        calculated at an applied flux equals to 
        $\Phi_c(D_1/D_2)$, as functions of 
        $D_1/D_2$. In both curves $R = 3 \xi_1$.}
        \label{fig:fig6}
    \end{figure}
    
    The behavior of the curve 
    resumes the previously described pattern for the 
    $R = 5 \xi_1$ case only after reaching a minimum in this flux.
    This occurs due the relationship between the critical 
    flux for the formation of a phase-slip $\Phi_c$ and 
    the "upper critical flux of the second band" $\Phi_{uc}$, 
    defined in the previous Section.
    As shown in the main panel of Fig.~\ref{fig:fig6}, 
    $\Phi_c$ is larger than $\Phi_{uc}$ for small values of 
    the ratio $D_1/D_2$. As a consequence, at the stability limit, the second band is superconducting due only to the coupling with the first band. 
    As presented in the inset of 
    Fig.~\ref{fig:fig6}, this 
    implies that for these values of $D_1/D_2$, 
    a very small portion of current is carried by the 
    second band. As we have seen in the previous 
    discussion, this means that the stronger 
    band dominates the process of phase-slip 
    formation and results in a larger critical flux. 
    Increasing $D_1/D_2$, the curves 
    $\Phi_c$ and $\Phi_{uc}$ intersect, activating the second band. 
    This leads to a recurrence of the behavior observed for $R = 5\xi_1$.
    As we can see then, the minimum in $\Phi_c$ originates 
    from the transition of the weaker band from passive 
    to active. This behavior is absent is the curve for 
    $R = 5\xi_1$ in Fig.~\ref{fig:fig4} because 
    $\Phi_{uc} > \Phi_c$ in the entire region of $D_1/D_2$ 
    displayed in this figure.

\begin{figure}[t!]
        \centering
        \includegraphics[width=0.48\textwidth]{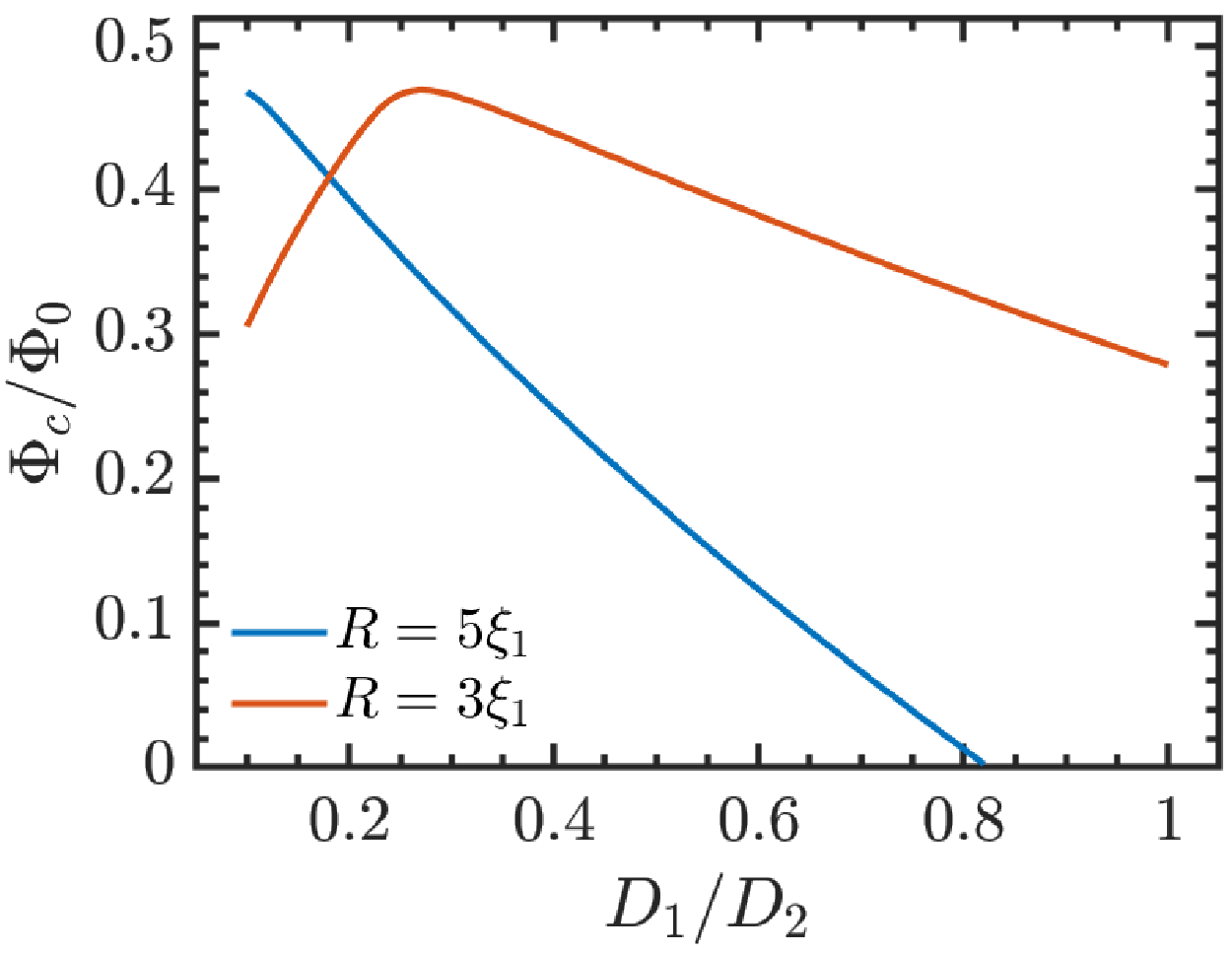}
        \caption{(Color online) Critical flux for the transition $n_i = 1 \rightarrow n_f = 0$ as a function of $D_1/D2$.}
        \label{fig:fig11}
    \end{figure}

    Our method is also capable of calculating the critical flux for the inverse transition \textit{i.e.} between an initial state winding number equals to $1$ in and a final state with winding number equals to $0$.
    When the external flux is being swept down, $\Phi_c$ for such transition is defined as the first flux at which the initial state with $n_i = 1$ becomes unstable against a perturbation with $n_f = 0$.
    In Fig.~\ref{fig:fig11}, this $\Phi_c$ as a function of $D_1/D_2$ is shown for $R = 5\xi_1$ (blue curve) and $R = 3\xi_1$ (red curve). As can be seen, for $R = 5\xi_1$ the critical flux monotonically decreases with $D_1/D_2$ and at a certain point reaches $0$. This means that is necessary to invert the flux sign to get an unstable $n_i = 1$. The ring with $R = 3\xi_1$ behaves at large $D_1/D_2$, but present the same non-monotonically behavior discussed in Fig.~\ref{fig:fig4} at small $D1/D_2$.

    \subsection{Critical flux dependence on the bath temperature}
    
\begin{figure}[t!]
        \centering
        \includegraphics[width=0.48\textwidth]{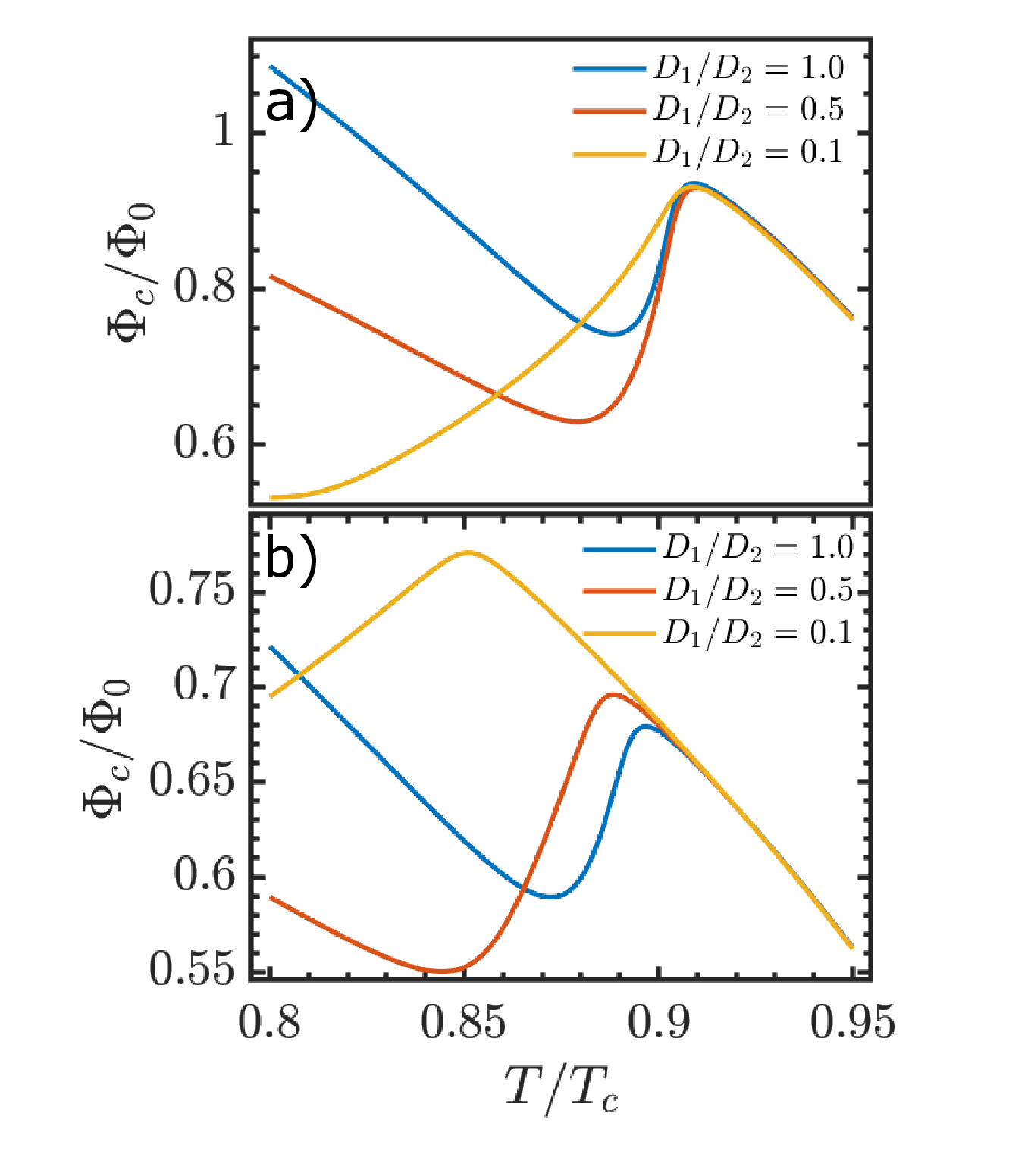}
        \caption{(Color online) The critical flux as a 
        function of temperature for $R = 5\xi_1$ (panel $a)$) 
        and $R = 3\xi_1$ (panel $b)$). In each panel, 
        we show the critical flux for $D_1/D_2 = 1.0$, $0.5$ 
        and $0.1$ (blue, red and yellow curves, respectively).}
        \label{fig:fig7}
    \end{figure}

    Let us now investigate how the critical flux 
    behaves 
    with temperature. Fig.~\ref{fig:fig7} 
    depicts $\Phi_c$ for three different values of 
    $D_1/D_2$ with temperature ranging from $T = 0.8 T_c$ 
    to $T = 0.95 T_c$. Panel $a)$ shows the critical 
    flux for $D_1/D_2 = 1.0$ (blue curve), $D_1/D_2 = 0.5$ 
    (red curve) and $D_1/D_2 = 0.1$ (yellow curve).
    
    As we can see, the blue 
    and red curves exhibit similar behavior; 
    namely, for small values of $T$, the critical flux 
    decreases as the temperature increases. 
    This is the expected behavior in a 
    superconducting system. On the other hand, the 
    yellow curve presents the intriguing behavior 
    of an increasing critical flux with temperature value. 
    This occurs because, for such small values 
    of $D_1/D_2$, the weaker band is passive throughout 
    this temperature region. As the amount of 
    supercurrent 
    carried by the weaker condensate 
    decreases, the system becomes more dominated 
    by the stronger condensate, leading to an 
    increase in the critical flux. This behavior 
    lasts until superconductivity in the weaker 
    band is completely destroyed, after which the 
    critical flux decreases with temperature, 
    a typical 
    behavior of single band superconductors. 
    As can be seen, the blue and red curves also 
    display regions where $\Phi_c$ increases with 
    the temperature. The mechanism behind this 
    is exactly the same one detailed for the 
    yellow curve. Finally, we note that 
    superconducting rings with radius $3\xi_1$ 
    depicted in panel $b)$ presents the same 
    qualitative behavior of the yellow curve. 
    In this case, the temperature values where 
    the maximum and minimum of the critical flux 
    occurs are smaller than the ones obtained in 
    panel $a)$. As we have discussed above, this 
    occurs because a smaller radius makes the 
    suppression of superconductivity easier 
    for both condensates.
    
    \subsection{The phase soliton state}
     
    In order to maintain the previous discussion
    as straightforward as possible, up to this point we have not described
    how the final state varies as we change the value $D_1/D_2$. In other words, we have not investigated what the winding numbers are in the final state for the various values of $D_1/D_2$
    presented in Fig.~\ref{fig:fig4}.
    
    Given the small radii and the relatively high temperature
    employed in the cases of Fig.~\ref{fig:fig4},
    the critical flux we found is never larger than $1.5 \Phi_0$. 
    This  means that winding numbers larger than $1$ are thermodynamically unfavorable. Nevertheless, as stated earlier,
    it is still possible to obtain a state with a phase-slip
    in only one of the bands.
    
    For $R = 5 \xi_1$ ($R = 3 \xi_1$), for instance, 
    $D_1/D_2 \geq 0.8$ ($D_1/D_2 \geq 1.1$)
    the system always transits from the Meissner state
    ($\Delta_j = a_j^{MS}$) to a phase-locked state
    ($\Delta_j = a_j^{PS}e^{ikx}$), where $k = 1/R$
    for both bands. In other words, 
    both bands display
    the same winding number. For $D_1/D_2 < 0.8$ ($D_1/D_2 < 1.1$)
    the Meissner state may be succeeded by a less trivial state
    with a different phase winding at each band,
    which is known as a phase soliton. The formation 
    or not of the phase soliton depends on the exact 
    value of the applied flux. For instance, if the 
    flux is much larger than the critical flux, the 
    system will go to a phase-locked state, even 
    if its diffusion coefficient ratio allows for 
    the formation of a soliton.
    
    \begin{figure}[t!]
        \centering
        \includegraphics[width=0.48\textwidth]{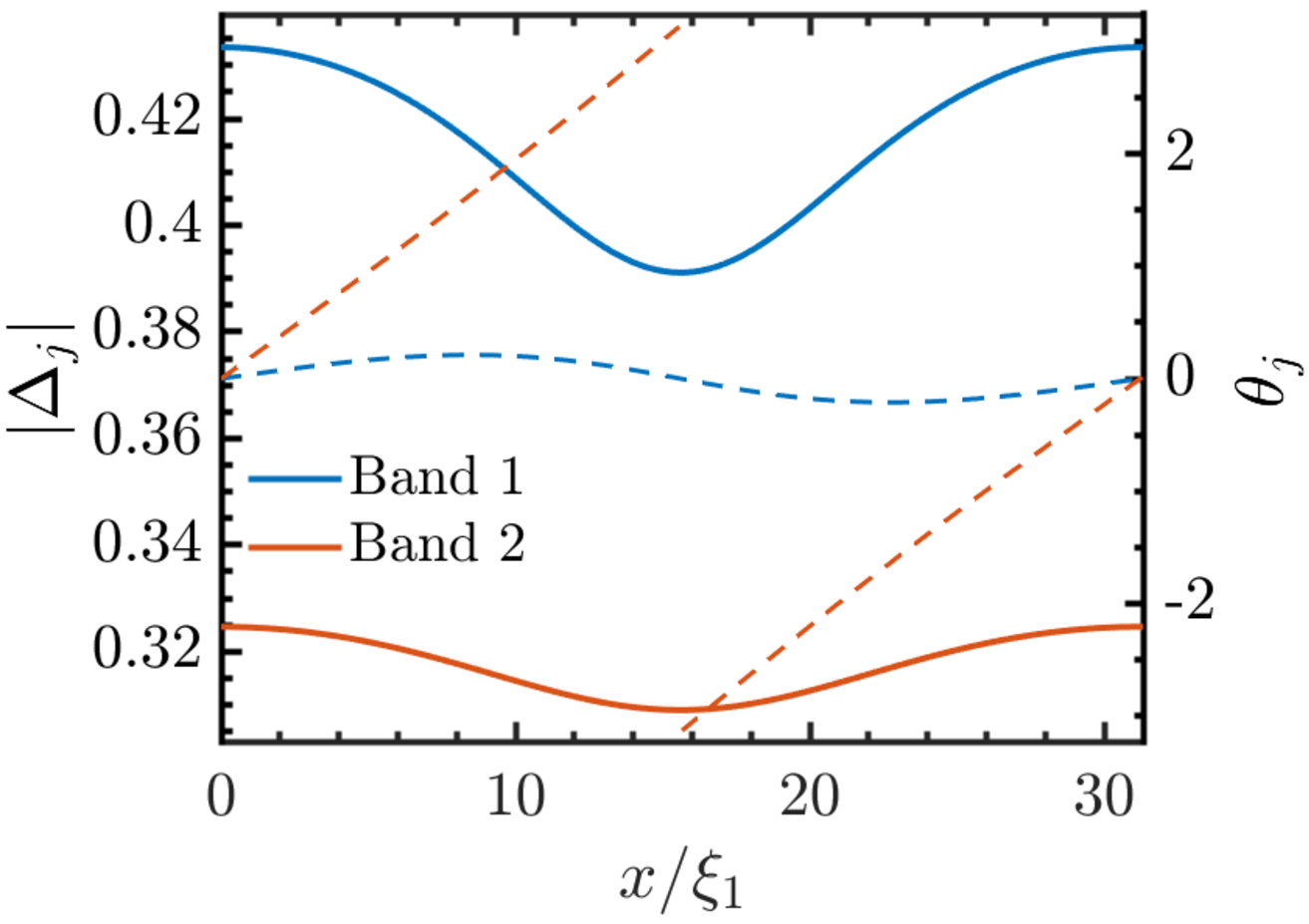}
        \caption{(Color online) The spatial dependence of the magnitudes (solid lines) and phases (dashed lines) of the order parameters correspondent to the first (blue) and second (red) band in a soliton state. The left and the right vertical axes give the magnitudes and the phases, respectively. Here, $R = 5\xi_1$, $D_1/D_2 = 0.5$ and the applied flux is the critical flux for this set of parameters $\Phi_c = 0.82\Phi_0$.}
        \label{fig:fig8}
    \end{figure}
    
    On the other hand, if the applied flux is close 
    enough to $\Phi_c$ in this region of diffusion 
    coefficient ratio, the system transits from the 
    Meissner state to the a phase soliton state.
    In Fig.~\ref{fig:fig8} we show the spatial 
    dependence of the magnitude of the order 
    parameters and their phases in a typical 
    soliton state. The parameters used are 
    $R = 5 \xi_1$, $D_1/D_2 = 0.5$, with an 
    applied flux slightly above the critical flux 
    obtained for these parameters, $\Phi_c = 0.82 \Phi_0$.
    Unfortunately, our semi-analytical method is 
    not able to predict whether the final state will be phase-locked 
    or the phase soliton state. We note that, even though $n_1 = 0$ and $n_2 = 1$ in the final state, $\Phi_c$ is obtained from a initial perturbation with winding number equals to $1$ fro both bands.

\begin{figure}[t!]
        \centering
        \includegraphics[width=0.48\textwidth]{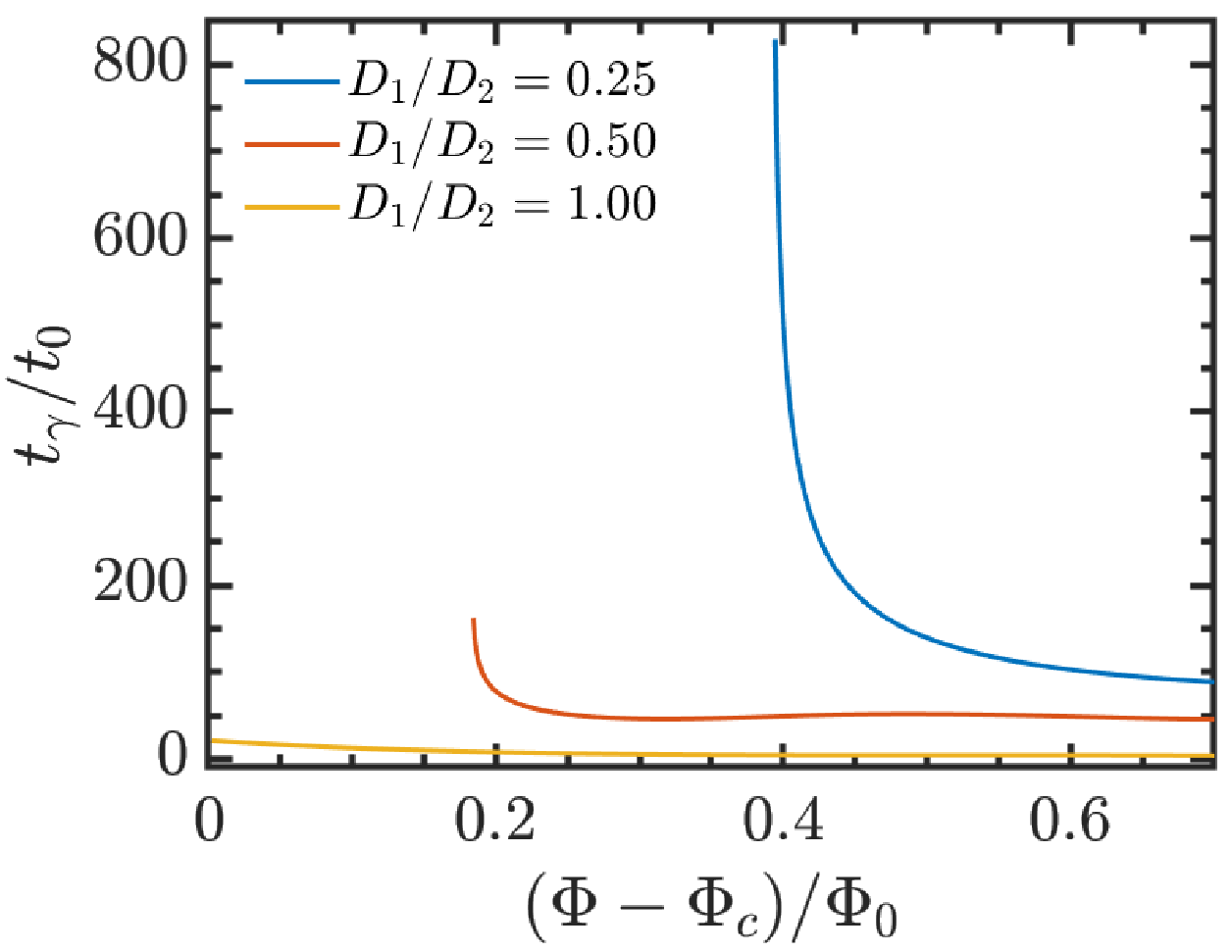}
        \caption{(Color online) $t_{\gamma}$ as a function of $(\Phi-\Phi_c)$ for $D_1/D_2 = 0.25$ (blue curve), $0.50$ (red curve) and $1.00$ (yellow curve). Here $R = 5\xi_1$ and $T = 0.8T_c$.}
        \label{fig:fig10}
    \end{figure}

    The possible stabilization of a phase soliton (as a metastable state) can be characterized by the time 
    difference between the formation of phase-slips in each 
    condensate, hereafter referred as $t_{\gamma}$. For 
    states with $n_1 = n_2$, $t_{\gamma}$ is finite, while 
    for the soliton state, where the phase-slip never 
    occurs in one of the order parameters, $t_{\gamma} 
    \rightarrow \infty$. We now investigate how $t_{\gamma}$ depends on $D_1/D_2$ and on the applied flux. To do this, we numerically solve the 
    Ginzburg-Landau equations for different flux values, always initializing the 
    system with $n_1 = n_2 = 0$. The results are shown in Fig.~\ref{fig:fig10}, where we present $t_{\gamma}$ as a function of $\Phi-Phi_c$ (with $\Phi_c$ being the critical flux for correspondent value of $D_1/D_2$) in order to compare different values of $D_1/D_2$ on equal footing. As can be seen, as $\Phi$ gets closer to $\Phi_c$, $t_{\gamma}$ varies 
    slowly for $D_1/D_2 = 1.00$, but presents a sharp increase that 
    eventually leads to a divergence at an appropriately small $\Phi$ for $D_1/D_2 = 0.25$ and $0.50$. This 
    is in agreement with our previous result that a phase 
    soliton is only possible for $D_1/D_2 < 0.8$ 
    and a phase soliton is only possible for 
    parameters that result in $t_{\gamma} \rightarrow \infty$.

\begin{figure}[t!]
        \centering
        \includegraphics[width=0.48\textwidth]{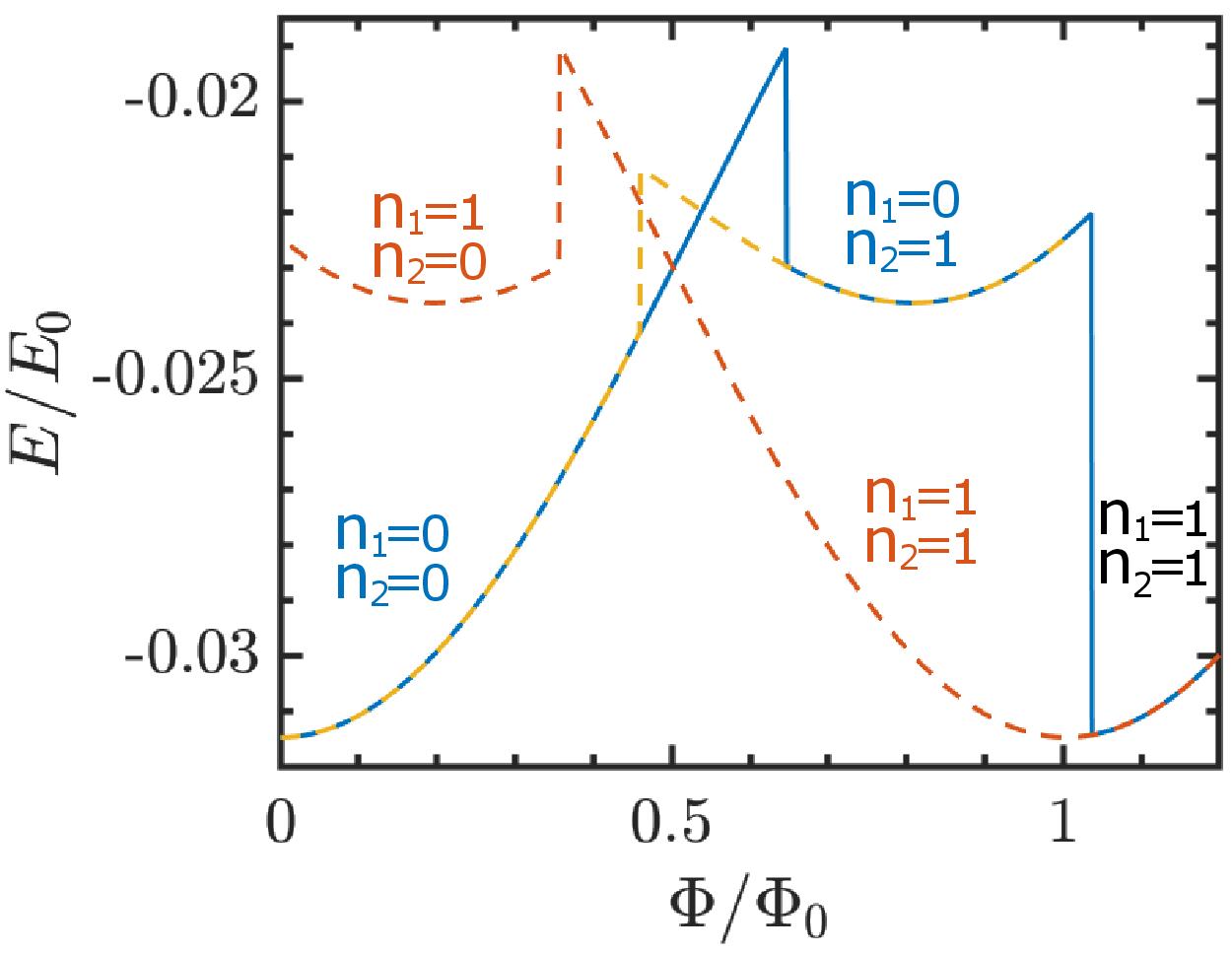}
        \caption{(Color online) Energy as a function of the applied flux for $R = 5\xi_1$ and $D_1/D_2 = 0.25$. The numbers $n_1$ and $n_2$ denotes the winding number of each order parameter. Solid blue and dashed red lines depict the regimes where the flux is being swept up and down in the range $0 \leq \Phi \leq 1.2\Phi_0$, respectively. In the dashed yellow curve, the flux is decreased from $\Phi = \Phi_0$ down to $\Phi = 0$.}
        \label{fig:fig9}
    \end{figure}

    Let us now investigate the stability of our soliton state for different values of applied flux. To do so, we start with the system in a state with no flux threading the superconducting loop and with winding number zero in both order parameters, and gradually change the applied flux in units of $\Delta\Phi = 0.001\Phi_0$. For each value of the applied flux we let the system relax to its equilibrium state. In Fig.~~\ref{fig:fig9} we show the system energy as a function of the applied flux in the range $0 \leq \Phi \leq 1.2\Phi_0$, for a superconducting ring with $R = 5\xi_1$ and $D_1/D_2 = 0.25$. Solid blue and dashed red lines correspond to the regimes where the flux is being swept up and down, respectively. The expression for the free energy density of our system can be found in Appendix \ref{sec:appB}.

    As can be seen from the blue curve, the system stays at the zero winding number state up to the critical flux $\Phi_c = 0.647\Phi_0$, as predicted from our semi-analytical method, and transits to a phase soliton state. During the transition a phase-slip occurs only in the second order parameter. The soliton state remains stable up to $\Phi = 1.037\Phi_0$. As we can see, the soliton state can be stable for a large window of the applied flux values, making it important to the behavior of the system. We note, though, that such window depends on the rate at which the applied flux is being increased, decreasing if the flux step is too large. If we now decrease the applied flux, as shown in the red curve, we see that the state $n_1 = n_2 = 1$ is stable down to $\Phi = 0.356\Phi_0$, where it transits to a different soliton state, with $n_1 = 1$ and $n_2 = 0$, which still remains stable at zero applied flux.

    In the dashed yellow curve, on the other hand, we start from the soliton state at $Phi=Phi_0$ and the external flux is decreased. In this case, when we decrease the flux, the system transits from the soliton state $n_1 = 0$ and $n_2 = 1$ to the state $n_1 = n_2 = 0$ at $\Phi = 0.459\Phi_0$. From the comparison of the different curves in Fig.~~\ref{fig:fig9} it becomes clear that a soliton state can be metastable for a large range of applied flux, with its occurrence being highly dependent on the system history.
    
\section{\label{sec:level5}Conclusions}

In the present work, 
we have studied the process of phase-slip mediated transitions in 
two-band superconducting rings under an applied flux. 
By combining linear stability theory with the time 
dependent Ginzburg-Landau equations for superconducting 
systems with two condensates, we have developed a 
semi-analytical method to determine the critical 
flux above which a transition must occur. 
We have validated our new method by comparing it 
with the results of the numerical solution of the Ginzburg-Landau equations. As we show, 
the agreement between the two methods is excellent.

Having verified the validity of our model, we used 
it to study how the behavior of our system depends 
on the diffusion coefficient ratio of each band and 
the temperature. A number of distinct features 
were shown, such as the unexpected increase of the 
critical flux with temperature for certain system 
parameters. Another unique behavior of such systems 
is the possibility of a phase soliton state, which displays order parameters with different winding numbers. 
We note that, with a suitable modification 
in the parameters, our model could be applied to a 
bilayer composed of two single-band superconductors.
We note that these characteristic physical phenomena can also be investigate in superconducting layered rings, as shown in Ref.~\onlinecite{bluhm2006}. In such a case, an effectively two order parameters system emerges due to the coupling between the order parameter of each layer. In particular, the non-monotonic behavior of the critical flux with temperature can be accomplished in bilayer ring composed of different superconducting materials.

\begin{acknowledgments}
LRC and ES thank the Brazilan Agency Fundação de Amparo à Pesquisa do Estado de São Paulo (FAPESP) for financial support (process numbers 20/03947-2, 20/10058-0 and 12/04388-0). DD acknowledges support from CNEA, CONICET , ANPCyT ( PICT2019-0654) and  UNCuyo (06/C026-T1).
\end{acknowledgments}

\appendix

\section{\label{sec:appA}Coefficients $M_{ij}$}

Substituting Eq.\ref{eqn:eqn11b} in Eq.\ref{eqn:eqn11}, we have for the first order parameter and specific value of $n_f$:

\begin{eqnarray}
   \lambda_{n_f}\hat{a}_{n_f}^1 = &-&(n_f/R-A)^2\hat{a}_{n_f}^1+\left(\chi_1-2a_1^2\right)\hat{a}_{n_f}^1 \nonumber \\
   &-&a_1^2\hat{a}_{-n_f}^{1*}+\gamma\hat{a}_{n_f}^2 \;, \label{eqn:A1}
\end{eqnarray}
and:
\begin{eqnarray}
   \lambda_{-n_f}\hat{a}_{-n_f}^{1*} = &-&(-n_f/R+2n_i/R-A)^2\hat{a}_{-n_f}^{1*} \nonumber \\
   &+&\left(\chi_1-2a_1^2\right)\hat{a}_{-n_f}^{1*}
   -a_1^2\hat{a}_{n_f}+\gamma\hat{a}_{-n_f}^{2*} \;, \label{eqn:A2}
\end{eqnarray}

From Eq.~\ref{eqn:A1}, we can identify that the coefficients $M_{11}$, $M_{12}$, $M_{13}$ and $M_{14}$ are equal to the terms multiplying $\hat{a}_{n_f}^1$, $\hat{a}_{-n_f}^1$, $\hat{a}_{-n_f}^2$ and $\hat{a}_{n_f}^2$, respectively. The same follows for Eq.~\ref{eqn:A2} and the analogous equations for the second order parameter.

The coefficients $M_{ij}$ in Eq.~(\ref{eqn:eqn14}) can then be written as:
\begin{widetext}
    \begin{eqnarray}
        M_{11} &=& -(n_f/R-A)^2+(\chi_1-2a_1^2)\;, \nonumber \\
        M_{12} &=& -a_1^2\;, \nonumber \\
        M_{13} &=& 0\;,
        \nonumber \\
        M_{14} &=& \gamma\;,
        \nonumber \\
        M_{21} &=& -a_1^2\;, \nonumber \\
        M_{22} &=& -(-n_f/R+2n_i/R-A)^2+(\chi_1-2a_1^2)\;,
        \nonumber \\
        M_{23} &=& \gamma\;,
        \nonumber \\
        M_{24} &=& 0\;,
        \nonumber \\
        M_{31} &=& 0\;,
        \nonumber \\
        M_{32} &=& \frac{\eta_1}{\eta_2}\gamma\;,
        \nonumber \\
        M_{33} &=& -\frac{D_2}{D_1}(-n_f/R+2n_i/R-A)^2+(\chi_2-2a_2^2)\;, \nonumber
        \nonumber \\
        M_{34} &=& -a_2^2\;,
        \nonumber \\
        M_{41} &=& \frac{\eta_1}{\eta_2}\gamma\;,
        \nonumber \\
        M_{42} &=& 0\;,
        \nonumber \\
        M_{43} &=& -a_2^2\;, 
        \nonumber \\
        M_{44} &=& -\frac{D_2}{D_1}(n_f/R-A)^2+(\chi_2-2a_2^2)\;.
    \end{eqnarray}
\end{widetext}

\section{\label{sec:appB}System free energy}

The free energy density of our system can be expressed as: \cite{vargunin2020}

\begin{eqnarray}
    F &=& \left(-\chi_1+\frac{1}{2}|\Delta_1|^2\right)|\Delta_1|^2+\frac{\eta_2}{\eta_1}\left(-\chi_2+\frac{1}{2}|\Delta_2|^2\right)|\Delta_2|^2 \nonumber \\
    &&+|\left (-i\mbox{\boldmath $\nabla$}-{\bf A}\right )\Delta_1|^2+\frac{\eta_2D_2}{\eta_1D_1}|\left (-i\mbox{\boldmath $\nabla$}-{\bf A}\right )\Delta_2|^2 \nonumber \\
    &&+\gamma\left(\Delta_1\Delta_2^*+\Delta_1^*\Delta_2\right)
\end{eqnarray}

In an equilibrium situation, we can take Eqs.~\ref{eqn:eqn6} with $\partial\Delta_j/\partial t = 0$ and $\varphi = 0$. Multiplying by $\Delta_j$ and integrating by parts, one can obtain a simple expression for the free energy density:

\begin{equation}
    F = -\frac{1}{2}|\Delta_1|^4-\frac{1}{2}\frac{\eta_2}{\eta_1}|\Delta_2|^4
\end{equation}

\bibliography{references}

\end{document}